\documentclass[aoas,MSNbibl,nameyear,dvips]{arximspdf}
\usepackage{algorithm}
\usepackage{algorithmic}
\usepackage{multirow}
\usepackage{dcolumn}
\usepackage{graphicx}

\doi{10.1214/11-AOAS486}
\volume{5}
\issue{4}
\pubyear{2011}
\firstpage{2603}
\lastpage{2629}

\makeatletter

\def\bsuffix #1{#1}

\newcolumntype{d}[1]{D{.}{.}{#1}}

\newcommand{\cov}{\operatorname{Cov}}
\newcommand{\e}{\mathrm{E}}
\newcommand{\pmse}{\operatorname{PMSE}}
\newcommand{\rmsd}{\operatorname{RMSD}}
\newcommand{\N}{\mathrm{N}}

\newcommand{\bh}{\mathbf{h}}
\newcommand{\bu}{\mathbf{u}}
\newcommand{\bw}{\mathbf{w}}
\newcommand{\bx}{\mathbf{x}}
\newcommand{\by}{\mathbf{y}}
\newcommand{\bz}{\mathbf{z}}

\newcommand{\bI}{\mathbf{I}}
\newcommand{\bZ}{\mathbf{Z}}

\newcommand{\sbx}{\mathbf{x}}
\newcommand{\sby}{\mathbf{y}}

\newcommand{\bone}{\mathbf{1}}
\newcommand{\bzero}{\mathbf{0}}

\newcommand{\btheta}{\bolds\theta}
\newcommand{\blambda}{\bolds\lambda}
\newcommand{\bphi}{\bolds\phi}
\newcommand{\bGamma}{\bolds\Gamma}
\newcommand{\bSigma}{\bolds\Sigma}
\newcommand{\bgamma}{\bolds\gamma}
\newcommand{\bsigma}{\bolds\sigma}

\newcommand{\bbbr}{\mathbb{R}}
\newcommand{\bbbn}{\mathbb{N}}

\newcommand{\rmA}{\mathrm{A}}
\newcommand{\rmB}{\mathrm{B}}
\newcommand{\rmM}{\mathrm{M}}
\newcommand{\rmfg}{\mathrm{fg}}
\newcommand{\rmAB}{\mathrm{AB}}
\newcommand{\rmtrue}{\mathrm{true}}
\newcommand{\rmcont}{\mathrm{cont}}
\newcommand{\rmmean}{\mathrm{mean}}
\newcommand{\rmpool}{\mathrm{pool}}
\newcommand{\rmMAP}{\mathrm{MAP}}

\makeatother

\begin{document}
\begin{frontmatter}

\title{Bayesian matching of unlabeled marked point sets using random
fields, with an application to molecular alignment\thanksref{T1}}
\runtitle{Alignment of unlabeled marked point sets}

\thankstext{T1}{Supported by an EPSRC/University of Nottingham
studentship and a Leverhulme Research Fellowship.}

\begin{aug}
\author[A]{\fnms{Irina} \snm{Czogiel}},
\author[B]{\fnms{Ian L.} \snm{Dryden}\corref{}\ead[label=e1]{ian.dryden@nottingham.ac.uk }} and
\author[C]{\fnms{Christopher J.} \snm{Brignell}}
\runauthor{I. Czogiel, I. L. Dryden and C. J. Brignell}
\affiliation{Max Planck Institute for Molecular Genetics,
University of South Carolina and University of Nottingham}
\address[A]{I. Czogiel\\
Max Planck Institute\\
\quad for Molecular Genetics\\
Ihnestrasse 63-73\\
14195 Berlin\\
Germany} 
\address[B]{I. L. Dryden\\
Department of Statistics\\
LeConte College\\
University of South Carolina\\
Columbia, South Carolina 29208\\
USA}
\address[C]{C. J. Brignell\\
School of Mathematical Sciences\\
University of Nottingham\\
University Park\\
Nottingham\\
NG7 2RD\\
United Kingdom}
\end{aug}

\received{\smonth{10} \syear{2010}}
\revised{\smonth{5} \syear{2011}}

%
\begin{abstract}
Statistical methodology is proposed for comparing unlabeled mar\-ked
point sets, with an application to aligning steroid molecules in
chemoinformatics. Methods from statistical shape analysis are combined
with techniques for predicting random fields in spatial statistics in
order to define a suitable measure of similarity between two marked
point sets. Bayesian modeling of the predicted field overlap between
pairs of point sets is proposed, and posterior inference of the
alignment is carried out using Markov chain Monte Carlo simulation. By
representing the fields in reproducing kernel Hilbert spaces, the
degree of overlap can be computed without expensive numerical
integration. Superimposing entire fields rather than the configuration
matrices of point coordinates thereby avoids the problem that there is
usually no clear one-to-one correspondence between the points. In
addition, mask parameters are introduced in the model, so that partial
matching of the marked point sets can be carried out. We also propose
an adaptation of the generalized Procrustes analysis algorithm for the
simultaneous alignment of multiple point sets. The methodology is
illustrated with a simulation study and then applied to a data set of
31 steroid molecules, where the relationship between shape and binding
activity to the corticosteroid binding globulin receptor is explored.
\end{abstract}

%

\end{frontmatter}

\section{Introduction}
In many application areas it is of interest to compare marked point
sets, where measurements (marks) are available at various point
locations, and often the configurations of points are unlabeled in the
sense that there is no natural correspondence between the points in
each configuration. The task of comparing unlabeled marked point sets
has been of recent interest in statistical shape analysis, for example,
\citet{Greemard06}, \citet{Drydenetal07} and \citet
{Schmidler07}. As
opposed to these previous approaches, our method does not aim to model
point correspondences. Instead, the objects are compared by assuming a~common underlying reference field which gives rise to the spatial
distribution of the marks.

One example where the alignment of unlabeled marked point sets is of
practical importance comes from the fields of structural bioinformatics
and chemoinformatics, where
it is of great interest to align molecules. However, the
task is often very difficult.
The motivating application in
this paper is a data set comprising 31 steroid molecules which bind
to the corticosteroid binding globulin (CBG) receptor. For each
molecule, the Cartesian coordinates of
the atom positions, as well as the associated van der Waals radii, and
the partial atomic charge values at the atom positions are provided.
Here the marks at each point (atom) are either the van der Waals radii
or the partial charges.
The steroids fall into three activity classes with respect to their
binding activity to the CBG receptor [\citet{GoodSoRichards1993}],
and the main objective in this application is to compare the
molecules in order to obtain the common features in each of the
three groups and to examine whether these features are associated
with the type of binding activity.

We consider a simple model under which spatial
prediction of a reference field is carried out using the observed marks
in each configuration.
A measure of similarity between the two predicted fields is
then used to describe the similarity, taking into account
an unknown transformation between the point sets which gave rise to the
actually observed point coordinates. The parsimonious model
does not attempt to model accurately all aspects of the molecules in
our application. It is rather used to develop a Bayesian algorithm
based on Markov chain Monte Carlo (MCMC) simulations for matching,
which is demonstrated to work well
in our applications.
In this setting it is also possible to
introduce additional parameters (mask vectors) which allow for the
fact that only part of the point sets may be similar. By determining
and aligning only the similar parts of the given point sets, a~meaningful comparison can be carried out.

In Section \ref{secMolSimGeo} we motivate and describe our newly
developed measure of similarity
for comparing unlabeled marked point sets. The Bayesian
framework for the pairwise alignment and similarity
calculation is introduced in Section \ref{secparwise}. An extension
of this methodology to the simultaneous alignment of multiple
point sets is described in Section \ref{secMultiple}.
In Section \ref{secSimulation} we carry out
simulation
studies in two and three dimensions to validate our method. In Section
\ref{secApplication} we apply our methods to the steroids data and
assess the results with respect to their chemical relevance.
Finally, Section \ref{secdiscussion} concludes the paper with a
discussion.

\section{Similarity measures using spatial prediction} \label{secMolSimGeo}

\subsection{Random field model}
The starting point for our model is an underlying reference random
field $\{Z(\bx)\dvtx\bx\in\bbbr^m\}$ which is assumed to be
second-order stationary with a constant mean $E(Z(\bx))=\mu$ and a
positive definite covariance function
$\sigma(\bh)=\cov(Z(\bx),Z(\bx+ \bh)) = \sigma( -\bh
)$. Consider two marked
point sets $A$ and $B$, say, which are $z$-values at point locations
in the field
given by $\bx'^{\rmA}_i \in\bbbr^m, i=1,\ldots,k_A$, and
$\bx'^{\rmB}_j \in\bbbr^m, j=1,\ldots,k_B$. In a
vector representation, the marked
point sets $A$ and $B$ can therefore be written as
\[
\bz^{\rmA} = \{ z^{\rmA}(\bx'^{\rmA}_1),\ldots
,z^{\rmA}(\bx'^{\rmA}_{k_{\rmA}} ) \} ,\qquad
\bz^{\rmB} = \{ z^{\rmB}(\bx'^{\rmB}_{1}),\ldots
,z^{\rmB}(\bx'^{\rmB}_{k_{\rmB}})\},
\]
respectively. Note that the relative position of $A$ and $B$ as given
by $\{\bx'^{\rmA}_1,\ldots,\allowbreak\bx'^{\rmA}_{k_{\rmA}}\}$ and $\{\bx
'^{\rmB}_1,\ldots,\bx'^{\rmB}_{k_{\rmB}}\}$ is
special because the spatial distribution of the marks within each point
set and also
the spatial distribution of the joint set of marks
$\bz^{\rmAB} = \{ z^{\rmA}(\bx'^{\rmA}_1),\ldots
,z^{\rmA}(\bx'^{\rmA}_{k_{\rmA}}), z^{\rmB}(\bx'^{\rmB
}_{1}),\ldots,z^{\rmB}(\bx'^{\rmB}_{k_{\rmB}}) \}$
directly\vspace*{1pt} reflect the properties of $\{Z(\bx)\dvtx\bx\in\bbbr
^m\}$. In that sense it can be regarded as the true relative position.

However, in real-life data sets the given marked point sets are often
provided in arbitrary locations, that is, before being recorded each
point set is transformed to new locations
$\bx^{\rmA}_i =
\Phi_{\rmA}(\bx'^{\rmA}_i),
i=1,\ldots,k_{\rmA}$, and $\bx^{\rmB}_j =
\Phi_{\rmB}(\bx'^{\rmB}_{j}),
j=1,\ldots,k_{\rmB}$, where $\Phi_{\rmA}\dvtx
\bbbr^m \to\bbbr^m$ and $\Phi_{\rmB}\dvtx\bbbr^m \to
\bbbr^m$ are unknown transformation functions
which are assumed to be 1--1 and onto. Hence, the inverse
transformations $\Phi_{\rmA}^{-1}$ and
$\Phi_{\rmB}^{-1}$ exist and satisfy
$\Phi_{\rmA}^{-1}\{ \Phi_{\rmA}(\bx) \} =
\bx
= \Phi_{\rmA} \{ \Phi_{\rmA}^{-1}(\bx) \}$ and
$\Phi_{\rmB}^{-1}\{ \Phi_{\rmB}(\bx) \} =
\bx
= \Phi_{\rmB} \{ \Phi_{\rmB}^{-1}(\bx) \}$,
respectively.

The basic inference problem we consider in this paper can
now be formulated as follows: if we are given the two marked point sets
$A$ and $B$ with $\bz^{\rmA}$ recorded at locations
$\{\bx^{\rmA}_1,\ldots,\bx^{\rmA}_{k_{\rmA}}
\}$ and $\bz^{\rmB}$ recorded at locations
$\{\bx^{\rmB}_1,\ldots,\bx^{\rmB}_{k_{\rmB}}
\}$, can we measure
how similar they are, taking into account the unknown transformation
$\Phi= \Phi_{\rmA} \Phi_{\rmB}^{-1}$ from
$B$ to $A$?
The method involves aligning the point sets by estimating the
transformation parameters in $\Phi$.


The particular choice of the set of potential transformations will
depend on the application. In our case the marked point sets
are the partial charges or the van der Waals radii of the steroid
molecules which are recorded in arbitrary positions and
orientations.
As steroid molecules in general are rigid (the word is derived from
``stereos''${}={}$``rigid'' in Greek), we consider the rigid body
transformations of translation and rotation, that is,
%
%
\begin{equation} \label{eqrigidbody}
\Phi(\bx) = \bGamma\bx+ \bgamma,\qquad
\bGamma\in
\operatorname{SO}(m), \bgamma\in\bbbr^m,
\end{equation}
where the space\vspace*{1pt} of special orthogonal matrices
$\operatorname{SO}(m)$ contains the rotation matrices which satisfy
$\bGamma^T\bGamma= \bGamma\bGamma^T = \bI_m$ and $|\bGamma|=1$. Other
more complicated transformations could be used, such as when more
dynamic aspects of molecule shape need to be taken into account. For
example, movement around rotatable bonds could be added if desired in
other applications. The choice of $\mu$ and $\sigma(\bh)$ in the random
field will also depend on the application.\looseness=1

In order to estimate the transformation parameters in $\Phi$, we
first consider predicting the underlying reference field $Z(\bx)$
using each point set separately. A similarity measure is then
defined which measures how close the two predicted fields are in a
certain relative position. Finally, we can estimate the unknown
transformation by maximizing the similarity measure or,
alternatively, by developing a statistical model based on the
similarity measure.

\subsection{Kriging}

In order to predict the underlying reference random field from each
point set, we consider simple kriging
[e.g., \citet{Cressie1993}, page~110] which assumes the mean
field $\mu= 0$. For the
steroid molecules with partial charge or van der Waals radius marks, it
makes sense to fix $\mu= 0$, so that a long way from the molecular
skeleton the predicted field is zero. We will use a sample variogram to
help suggest an appropriate covariance function.\looseness=1

Consider a general marked point set $\bz=\{ z(\bx
_1),\ldots,z(\bx_k)
\}$. If simple kriging is used to predict the value of the
underlying random field \mbox{$\{Z(\bx)\dvtx\bx\in\bbbr^m\}$} at a
location of interest $\bx_0$, say, a weighted average of the form
$\hat{Z}(\bx_0) = \sum_{i=1}^k u_i z(\bx_i) $ is sought so as to
minimize the prediction mean squared error $\pmse(\bu) = \e[
( \hat{Z}(\bx_0)-Z(\bx_0))^2 ]$ with respect to the
weight vector $\bu=(u_1,\allowbreak \ldots, u_k)^T$. Given the observed values
in $\bz$, the corresponding system of equations has
the solution
$\bu= \bSigma^{-1}\bsigma$, and the predicted
value for $Z(\bx_0)$
is given by
$\hat{Z}(\bx_0) = \bsigma(\bx_0)^{T} \bSigma
^{-1} \bz= \bu
^{T}\bz$,
where $\bsigma(\bx_0)=(\sigma(\bx_1-\bx
_0), \ldots,
\sigma(\bx_k-\bx_0) )^{T}$ and
$(\bSigma)_{ij}=\sigma(\bx_i-\bx_j)$, $1 \le i,j \le k$. For a
general location $\bx$
this yields the predicted field
%
%
\begin{equation}\label{eqKrigedField2}
\hat{Z}(\bx) = \bz^{T} \bSigma^{-1} \bsigma(\bx) = \sum_{i=1}^k
w_i \sigma(\bx_i-\bx) ,
\end{equation}
where the weight vector $\bw= (w_1,\ldots,w_k)^T = \bSigma^{-1}\bz$
is optimal in
terms of minimizing the PMSE if the underlying assumptions are met.
Note that
in some applications it may not be appropriate to assume $\mu=0$, in
which case one would work with the
mean corrected field
$Z(\bx)-\mu$, where $\mu$ is either known or estimated using
generalized least squares from each marked point set.

Using (\ref{eqKrigedField2}) and based on the observed data vectors
$\bz^{\rmA}$ and $\bz^{\rmB}$, we can
obtain a different prediction
of the underlying reference random field from each of the two marked
point sets $A$ and $B$, and
the resulting predicted fields $\hat{Z}_{\rmA}(\bx)$ and
$\hat{Z}_{\rmB}(\bx)$ then need to be compared.

\subsection{Function similarity and the Kernel Carbo index}
\label{subcomparison}
In
order to measure the similarity of the predicted fields $\hat
{Z}_{\rmA}(\bx)$ and
$\hat{Z}_{\rmB}(\bx)$, we require a~metric
space where the notion of
similarity can be defined by means of the
corresponding inner product. A commonly used metric space for
functions is the space of Lebesgue square-integrable functions
$L_2$, where the inner product has the form
%
%
\begin{equation}\label{L2}
\langle f,g \rangle_{L_2} = \int f(\bx) g(\bx) \,d \bx.
\end{equation}
%
Based on (\ref{L2}), an intuitive measure of similarity between two
functions $f$ and~$g$ can be formulated
which does not depend on the scales of $f$ and~$g$, that is,
%
\[
R_{\rmfg} = \frac{ \int f(\bx) g(\bx) \,d \bx}{ (
\int f(\bx)^2 \,d \bx)^{1/2} ( \int g(\bx)^2 \,d \bx
)^{1/2} }
= \frac{ \langle f,g \rangle_{L_2} }{ ( \langle f,f \rangle_{L_2}
\langle g,g \rangle_{L_2}
)^{1/2} } ,
\]
and so $R_{\rmfg}=1$ if $f = c g$, where $ c > 0$ is a
positive constant,
and $R_{\rmfg} = -1$ if $c < 0$. Note that $R_{\rmfg}$ is a
generalization of Pearson's correlation
coefficient for comparing two functions. Also note that, in general,
calculation of $R_{\rmfg}$ would involve numerical
integration over the domain, which may be
computationally demanding.

An alternative metric space for functions is a reproducing kernel
Hilbert space
(RKHS) that, for a
given reproducing kernel, 
can easily be constructed and is much simpler and quicker to use in practice.
This alternative is very useful for our model because the covariance function
$\sigma$ of the reference random field can be viewed as a reproducing
kernel on
$\bbbr^m \times\bbbr^m$
due to the properties of a general covariance function
(e.g., symmetric and positive definite). Hence, the corresponding
RKHS exists [\citet{Aronszajan1950}] and can be written as
$\mathcal{H}_{\sigma}=\{f | f( \bx)=\sum_{i=1}^{k_{\rmA}}
\alpha_i
\sigma(\bx_i^{\rmA}-\bx)\}$. In this space the
inner product of $f(
\bx)=\sum_{i=1}^{k_{\rmA}} \alpha_i \sigma( \bx
_i^{\rmA}-\bx) \in
\mathcal{H}_{\sigma}$ and $g( \bx)=\sum_{j=1}^{k_{\rmB}} \beta_j
\sigma(
\bx_j^{\rmB}-\bx) \in\mathcal{H}_{\sigma}$ has the form
\[
\langle f,g\rangle_{\mathcal{H}_{\sigma}}=\sum_{i=1}^{k_{\rmA}} \sum
_{j=1}^{k_{\rmB}} \alpha_i
\beta_j \sigma(\bx_i^{\rmA}-\bx_j^{\rmB}),
\]
which can be evaluated without
expensive numerical integration.

Note that we can view the kriging predictor (\ref{eqKrigedField2})
as a member of $\mathcal{H}_{\sigma}$, and, hence, we can use the RKHS
inner product $\langle\cdot,\cdot\rangle_{\mathcal{H}_{\sigma}}$ to
measure the
similarity between the predicted fields of $A$ and $B$.
Let $\hat{Z}_{\rmA}(\bx)=\sum_{i=1}^{k_{\rmA}}
w^{\rmA}_i \sigma(\bx_i^{\rmA}-\bx)$ and
$\hat{Z}_{\rmB}(\bx)=\sum_{j=1}^{k_{\rmB}}
w^{\rmB}_j \sigma(\Phi(\bx_j^{\rmB})-\bx
)$ denote the predicted fields of the marked point sets $A$ and $B$ in
the relative position defined by $\Phi=
\Phi_A \Phi_B^{-1}$.
The
similarity measure we propose in this paper has the form
%
%
\begin{eqnarray} \label{eqkernelcarbo}
C_{\rmAB}(\bphi)=\frac{\langle\hat{Z}_{\rmA}
,\hat{Z}_{\rmB} \rangle_{\mathcal{H}_{\sigma}}}{\|\hat
{Z}_{\rmA}\|_{\mathcal{H}_{\sigma}}\|\hat{Z}_{\rmB} \|_{\mathcal
{H}_{\sigma}}},
\end{eqnarray}
where
$\|\hat{Z}_{\rmM}\|^2_{\mathcal{H}_{\sigma}}= \langle\hat
{Z}_{\rmM} ,\hat{Z}_{\rmM} \rangle_{\mathcal
{H}_{\sigma}}$
$(M \in\{A,B\})$, and $\bphi$ denotes the parameter vector of the
unknown transformation $\Phi$.
The numerator
term measures the ``overlap'' of the fields (in a certain relative
position), whereas the denominator is a transformation invariant
normalizing constant
which
ensures that $C_{\rmAB}( \bphi) \in[-1,1]$.
Note that (\ref{eqkernelcarbo}) can also be interpreted as the cosine of
the angle between the two predicted fields in a certain relative
position.

We shall call the above similarity function the ``Kernel Carbo
function,'' as it is a modification of a similarity function proposed
by \citet{CarboLeydaArnau1980} in the context of field-based
molecular alignment. The fields considered in that original paper are the
electron densities of the two molecules under study, and the
similarity was defined in terms of the $L_2$ inner product given in
(\ref{L2}).
As both fields in our setting are members of the RKHS
$\mathcal{H}_{\sigma}$, the Carbo similarity function can be
``kernelized'' by replacing $\langle\cdot, \cdot\rangle_{L_2}$ with
$\langle\cdot,\cdot\rangle_{\mathcal{H}_{\sigma}}$, which has the
advantage that
calculating (\ref{eqkernelcarbo}) does not require evaluation of
overlap integrals over $\bbbr^m$ for any choice of
positive definite covariance function.

For the reproducing kernel we shall consider the istropic Mat{\'e}rn
covariance function,
where the covariance
of the field between any pair
of points~$\bx, \by$ is given by
%
%
\begin{equation} \label{eqmatern}
\sigma(\bx- \by) = \frac{1}{2^{\nu-1}\Gamma(\nu)}\biggl(\frac
{2\nu^{1/2}\|\bx-\by\|}{\rho}\biggr)^{\nu}
K_{\nu}\biggl(\frac{2\nu^{1/2}\|\bx-\by\|}{\rho}\biggr).
\end{equation}
This provides a flexible family of stationary covariance functions
[Stein (\citeyear{Stein99}), page 31].
With this particular parameterization [e.g., \citet
{HandcockWallis1994}], $\rho$
is a range parameter and $\nu$ determines the smoothness of the
random field. Moreover, $K_{\nu}(\cdot)$ is the modified Bessel function
of the third kind of order $\nu$ and $\Gamma(\cdot)$ is the Gamma
function.
Note that $\nu\to\infty$ corresponds to the Gaussian covariance function
%
%
\begin{equation}\label{Gaussian-kernel}
\sigma(\bx- \by)=
\exp\{-\|\sbx-\sby\|^2/\rho^2\} ,
\end{equation}
and in this particular case the $L_2$-Carbo index of our predicted fields
could be calculated analytically.


Optimizing (\ref{eqkernelcarbo})
with respect to the transformation
parameters yields the ``Kernel Carbo index''
%
%
\begin{equation} \label{eqOurSim}
C(A,B)=\sup_{\bphi} C_{\rmAB}(\bphi) = \sup_{\bphi} \frac{\langle\hat
{Z}_{\rmA} ,\hat{Z}_{\rmB} \rangle_{\mathcal
{H}_{\sigma}}}{\|\hat{Z}_{\rmA} \|_{\mathcal{H}_{\sigma
}}\|\hat{Z}_{\rmB} \|_{\mathcal{H}_{\sigma}}} ,
\end{equation}
in which configuration $B$ is transformed (by the relative
transformation function~$\Phi$)
to be as similar as possible to configuration $A$. In the case (\ref
{eqrigidbody}) where
the rigid body transformations in $\bbbr^m$ are considered, the
parameter vector $\bphi$ contains $m(m-1)/2$ Euler angles
for rotation and $m$ translation parameters, and in this case the
Kernel Carbo
index is invariant under the rigid body transformations of $A$ and
$B$.

Note that the optimization in (\ref{eqOurSim}) is not straightforward
in practice
due to local maxima. As an approximation to using the Kernel Carbo
index in~(\ref{eqOurSim}), we will therefore propose a Bayesian model
and find the value of the similarity index~(\ref{eqkernelcarbo}) at
the maximum a posteriori
(MAP) estimates of the transformation parameters. Also note that, in
situations where a dissimilarity rather than a similarity
measure is required, (\ref{eqkernelcarbo}) can be uniquely mapped
into the appropriate codomain using
%
%
\begin{equation} \label{eqdissimCarbo}
D_{\rmAB}(\bphi)=\frac{1-C_{\rmAB}(\bphi
)}{1+C_{\rmAB}(\bphi)} \in[0,\infty),
\end{equation}
and applying the same transformation to (\ref{eqOurSim}) or its MAP
equivalent then
yields a transformation invariant dissimilarity index between two marked
point sets.

\subsection{Masks}
In many applications it is of interest to match parts of objects rather
than the entire configurations. Our steroids application is one such
example because only a part of each molecule may fit into the binding
pocket of the common receptor and is hence relevant for the binding mechanism.
As a tool for matching only parts of the given configurations, we
consider a set of masks (indicator parameters) which signify if
individual points are included in the predicted field or not.
The masks therefore allow for the possibility that only parts of the
structures match, whereas other parts may have been generated by
different underlying reference fields or may be largely affected by
noise.

From now on we will just consider rigid body transformations
between the point sets, with rotation matrix $\bGamma$ and translation
vector $\bgamma$, although, as mentioned above, the
approach can
be extended to other transformations.

Let $\blambda_{\rmM}\,{=}\,(\lambda^{\rmM}_1,\ldots,
\lambda^{\rmM}_{k_M})^T$ be the mask vector for point set
$M$ ($M\,{\in}\,\{A, B\}$). Each
entry of the mask vector is an indicator function, that is,
$\lambda^{\rmM}_i\in\{0,1\}$ which determines
if the $i$th point of set $M$ is
considered to contribute to the matching parts
$(\lambda^{\rmM}_i=1)$ or not
$(\lambda^{\rmM}_i=0)$, $i=1,\ldots,k_M$. Taking the
mask vector into
account, the predicted version\vspace*{1pt} of the common reference
field~ba\-sed
on $M$ then has the form
\mbox{$\hat{Z}_{\rmM}(\bx;\blambda_{\rmM})=\sum
_{i\dvtx\lambda^{\rmM}_i=1}
w^{\rmM}_i(\blambda_{\rmM})\sigma(\bx
_i^{\rmM}-\bx)$},
and the\vspace*{-2pt} resulting partial Kernel Carbo function for two masked
fields $\hat{Z}_{\rmA}(\bx;\blambda_{\rmA})$
and $\hat{Z}_{\rmB}(\bx;\blambda_{\rmB})$
in a
certain relative position becomes
%
%
\begin{equation}\label{PK}
C_{\rmAB}(\bGamma,\bgamma,\blambda_{\rmA
},\blambda_{\rmB})=
\sum_{i\dvtx\lambda^{\rmA}_i=1}
\sum_{j\dvtx\lambda^{\rmB}_j=1}
\tilde{w}^{\rmA}_i(\blambda_{\rmA})
\tilde{w}^{\rmB}_j(\blambda_{\rmB})
\sigma\bigl(\bx_i^{\rmA}-(\bGamma
\bx_j^{\rmB}+\bgamma)\bigr),\hspace*{-30pt}
\end{equation}
where the tilde indicates
that the kriging weights are normalized by the corresponding term in
the normalizing constant, that is,
$\tilde{w}^{\rmM}_i(\blambda_{\rmM})=w^{\rmM}_i(\blambda_{\rmM
})/\allowbreak N_{\rmM}(\blambda_{\rmM})$,
with
$N_{\rmM}(\blambda_{\rmM})=\|\hat{Z}_{\rmM}(\bx;\blambda_{\rmM
})\|_{\mathcal{H}_{\sigma}}$.
The partial Kernel Carbo index can then be obtained by maximizing (\ref{PK})
over the transformation and mask parameters.

Optimizing
the similarity measure (\ref{PK}) over all possible subsets
is very challenging due to the
combinatorial nature of the search space.
Instead we use a Bayesian model
to obtain the MAP estimates of the similarity
transformations and
masks and then evaluate (\ref{PK}) at the MAP, which approximates
the maximization of (\ref{PK}). Rather than trying to develop a
realistic probabilistic model for the data, we therefore view
the Bayesian model and the resulting MCMC scheme as a practical
approach for
generating an algorithm to match two spatial point patterns. Also,
apart from transforming the problem into a more tractable one, the
Bayesian setting allows the introduction of prior information about the
parameters which will be useful, for example, to prevent excessive masking.

\section{Bayesian pairwise
alignment of marked point sets} \label{secparwise}
\subsection{Likelihood} \label{sublike}
With the assumption that the similar parts of the two point sets are
noisy pointwise observations of the same underlying reference
field, we define the likelihood for the two marked point sets
$ \bz^{\rmA} =\{z^{\rmA}(\bx^{\rmA}_1), \ldots,
z^{\rmA}(\bx^{\rmA}_{k_{\rmA}})\}$ and
$\bz^{\rmB} =\{z^{\rmB}(\bx^{\rmB}_1), \ldots,
z^{\rmB}(\bx^{\rmB}_{k_{\rmB}})\}$ in the relative
position defined by $\bGamma$ and $\bgamma$ as
%
%
\begin{equation} \label{eqlikelihood}
L(\bz^{\rmA}, \bz^{\rmB}
|\btheta
,\bgamma,\blambda_{\rmA},\blambda_{\rmB},\tau
) \propto\tau
\exp(-\tau D_{\rmAB}(\bGamma,\bgamma,\blambda
_{\rmA},\blambda_{\rmB})),
\end{equation}
where $\btheta$ denotes the vector of the Euler angles which
specifies a rotation matrix~$\bGamma(\btheta)$, $\bgamma$ denotes
a displacement vector between $A$ and $B$, $\tau\in\bbbr^+$ is a
precision parameter,
and
$D_{\mathrm{AB}}(\bGamma,\bgamma,\blambda_{\mathrm{A}},\blambda_{\mathrm{B}})$
is the dissimilarity function based on (\ref{eqdissimCarbo}) and (\ref{PK}).
Here, the mask vectors play a
similar role as the labeling matrices in \citet{Greemard06},
\citet{Drydenetal07} and \citet{Schmidler07}, except in our
framework there
is no need to establish
correspondences between points in $A$ and~$B$.
Instead, the mask vectors are defined separately for each point set.
The pairwise correspondence does not need be estimated because
all possible pairs of atoms are considered in the model, and the pairs
are weighted
according to how far apart they are during the matching.

Note that if $\tau$ is fixed, the likelihood is maximized at the same rotation,
translation and mask parameter estimates that give the maximum value of
the partial
Kriged Carbo index (\ref{PK}). This, and the fact that it performed well
in pilot simulations, provides the motivation for the
use of this likelihood. 
Other choices include the half-normal likelihood
\[
L(A,B |\btheta,\bgamma,\blambda_{\rmA
},\blambda
_{\rmB},\tau
) \propto\tau^{1/2}
\exp(-\tau D^2_{\rmAB}(\bGamma,\bgamma,\blambda
_{\rmA},\blambda_{\rmB})),
\]
which is less accommodating of outliers but might be preferable in some
situations.

\subsection{Prior distributions and posterior sampling}
\label{subpostsampling}
We do not have any prior information about the rigid body parameters
$\btheta$ and $\bgamma$ so that they are treated
as uniformly
distributed on $\operatorname{SO}(m)$ and on a large bounded region in
$\bbbr^m$,
respectively. The uniform distribution on $\operatorname{SO}(m)$ is
determined by the
probability measure which is invariant
under the group action. In the two-dimensional case, $f_U(\btheta)
\propto1$. For $m=3$, the appropriate density with respect to the
Lebesgue measure
depends on the parametrization of $\operatorname{SO}(3)$, and in this
paper we use the
Euler angles in the so-called $x$-convention where
%
%
\[
\bGamma(\btheta)=
\pmatrix{\cos\theta_3 \!&\! \sin\theta_3 \!&\! 0 \cr
- \sin\theta_3\! &\! \cos\theta_3 \!&\! 0 \cr
0 \!&\! 0 \!&\! 1\!}\!
\pmatrix{1 \!&\! 0 \!&\! 0 \cr
0 \!&\! \cos\theta_2 \!&\! \sin\theta_2 \cr
0 \!&\! - \sin\theta_2 \!&\! \cos\theta_2}\!
\pmatrix{\cos\theta_1 \!&\! \sin\theta_1 \!&\! 0 \cr
- \sin\theta_1 \!&\! \cos\theta_1 \!&\! 0 \cr
0 \!&\! 0 \!&\! 1 }.
\]
In that case, $f_U(\btheta) \propto\cos(\theta_2)$ and with the
domains $-\pi\leq\theta_1,\theta_3 < \pi$ and $-\pi/2 \leq\theta
_2 < \pi/2$,
every $\bGamma\in\operatorname{SO}(3)$ is uniquely determined apart
from a~singularity
at $\theta_2=-\pi/2$.

To prevent the situation where
only very few points are used in the field comparison, we
introduce a (fixed) penalty parameter $\zeta\ge0$ and a (fixed)
interaction parameter $\zeta_I \ge0$
to define the joint
prior density of the mask vectors as
\[
\pi(\blambda_{\rmA},\blambda_{\rmB}| \zeta
, \zeta_I) \propto
\zeta^{\sum_i \lambda_i^{\rmA}+\sum_i \lambda
_i^{\rmB}}
+ \zeta_I^{ \sum_{i \stackrel{\mathrm{A}}{\sim} j}
| \lambda_i^{\rmA}- \lambda_j^{\rmA} | +
\sum_{i \stackrel{\mathrm{B}}{\sim} j}
| \lambda_i^{\rmB}- \lambda_j^{\rmB} | } ,
\]
where\vspace*{1pt} $i \stackrel{\mathrm{M}}{\sim} j$ means that points $i$
and $j$ are neighbors within $M$ ($M \in\{A,B\}$), for example,
if $\| \bx^{\rmM}_i - \bx^{\rmM}_j \| <
\delta$. Note that the dimensions of $\blambda_{\rmA}
\in\{0,1\}^{k_{\rmA}}$
and $\blambda_{\rmB} \in\{0,1\}^{k_{\rmB}}$
are fixed.
The penalty parameter $\zeta$ inherently comprises prior
assumptions about the extent of the matching parts of $A$ and $B$, with
higher~$\zeta$ indicating more prior matching points.
Also, if the interaction parameter~$\zeta_I$ is strictly greater
than~1, this
indicates clustering so that nearby
points within a point set are expected to be included (or excluded)
together in the matching.
Thus, a large positive~$\zeta_I$ would be used when we wish to
encourage contiguous
regions to be included in the matching, although we shall use $\zeta_I
= 1$
in our applications.

With the further assumptions that the precision parameter is Gamma
distributed a priori, that is, $\tau\sim\Gamma(\alpha,
\beta)$,
and that all unknown parameters are independent a priori,
the joint posterior conditioned on the given data is
\begin{eqnarray*}
&&\pi(\btheta,\bgamma,\blambda_{\rmA},
\blambda_{\rmB},\tau|\bz^A,\bz^B,\alpha,\beta,\zeta
,\zeta_I) \\
&&\qquad\propto\tau^{\alpha}
\exp\bigl\{-\tau\bigl(D_{\rmAB}(\bGamma,\bgamma
,\blambda_{\rmA},\blambda_{\rmB})+\beta
\bigr)\bigr\}\cdot
\pi(\blambda_{\rmA},\blambda_{\rmB}|\zeta
, \zeta_I)
\cdot f_U(\btheta).
\end{eqnarray*}
Note that this can be regarded as a mixture model over
$\{0,1\}^{k_{\rmA}} \times\{0,1\}^{k_{\rmB}}$.

We use MCMC to
sample from the posterior distribution. The resulting point
estimates for the rigid body parameters and the mask vectors can
then be\vadjust{\goodbreak} substituted into
$D_{\rmAB}(\bGamma,\bgamma,\blambda_{\rmA
},\blambda_{\rmB})$
to yield a point estimate of the dissimilarity measure
%
%
\begin{equation} \label{eqplugdist}
\hat{D}_{\rmAB}=D_{\rmAB}(\hat{\bGamma
},\hat{\bgamma},\hat{\blambda}_{\rmA},\hat
{\blambda
}_{\rmB}).
\end{equation}
In the MCMC scheme, $\tau$ is updated with a Gibbs step. Updated
versions of the other parameters are obtained in four blocks, each
using a Metropolis--Hastings step. For the rigid body parameters, we
use independent normal proposals, and a
proposal distribution for the masks vectors
$\blambda_{\rmA}$ and $\blambda_{\rmB}$ can be
obtained by choosing an entry at random and then switching its value
from zero to one or vice versa.

The algorithm we use ensures that the defined Markov chain is
irreducible, aperiodic and positive recurrent, and, hence, after a
large number of iterations the simulated value is approximately
generated from the posterior distribution. Due to the symmetry of the proposal
distributions, convergence to and sampling from the limiting
distribution in practice thereby results in an approximate
stochastic minimization of the dissimilarity term, and this behavior
can be emphasized by choosing a prior distribution with a large mean
for $\tau$. In fact, if one is mainly interested in obtaining point
estimates of the model parameters which provide a good superposition,
simulated annealing
[\citet{KirkpatrickGelattVecchi1983}] can be employed so that the
MCMC algorithm simulates from $
\pi(\btheta,\bgamma,\blambda_{\rmA
},\blambda
_{\rmB},\tau|A,B,\alpha,\beta,\zeta,\zeta_I )^{1/T},
\label{simann} $ where $T > 0$ is slowly reduced deterministically.
\label{pgsiman}

As with any MCMC scheme for a complicated high-dimensional problem, there
is a possibility that the chain will become stuck in a local region of maximum
posterior probability, and our application is no exception. Hence, judicious
use of proposal distributions
is required to escape such regions, for example, the use of
occasional large proposal variances.

Note that the partial Kriged Carbo index and the Bayesian model are
symmetric in terms of which point set is denoted as $A$ and
which point set is denoted as $B$. However, for a practical
implementation one of the points sets is chosen as $B$ and transformed
to be as close as possible to the other point set $A$.
As our method is simulation based, slightly different estimates will be obtained
in matching $A$ to $B$ and then $B$ to $A$. Hence, in our application
we carry out
both matches and then
take an appropriately symmetrized average of the estimated distance measures,
for example, their geometric mean.

\subsection{Multiple alignment} \label{secMultiple}
In the multiple alignment problem, the objective is to
simultaneously superimpose a set of $n$ marked point sets $M_1,
\ldots, M_n$. Previous approaches to this problem include
\citet{Drydenetal07} and \citet{RuffieuxGreen2009}.
Here, we adapt the generalized Procrustes analysis (GPA) algorithm
for discrete landmark data [e.g., \citet{DrydenMardia1998}, page
90] to our field-based approach. In
the classical GPA context it is of interest to find an alignment of
the given objects which minimizes the sum of their pairwise
squared distances. A similar goodness-of-fit criterion for the multiple
superposition of $n$ predicted masked fields can be formulated in
terms of their overall similarity as
%
%
\begin{eqnarray} \label{eqmultiCarbo}
&&C(\btheta,\bgamma,\blambda) \nonumber\\
&&\qquad= \sum_{i=1}^{n-1}
\sum_{j=i+1}^n \biggl\{\sum_{l\dvtx\lambda^i_l=1} \sum_{l'\dvtx\lambda^j_{l'}=1}
\tilde{w}^{i}_l(\blambda_{i}) \tilde{w}^{j}_{l'}(\blambda_{j})
\sigma\bigl((\bGamma_i \bx_{l}^{i}+\bgamma_i)\\
&&\hspace*{213.5pt}{}-(\bGamma_j
\bx_{l'}^{j}+\bgamma_j)\bigr)\biggr\},\nonumber
\end{eqnarray}
where $\blambda^T=(\blambda_1^T, \ldots, \blambda^T_n) \in
\{ 0,1 \}^{\sum_{i} k_i}$, $\btheta^T=(\btheta_1^T, \ldots,
\btheta_n^T) \in\bbbr^{m(m-1)n/2}$ and $\bgamma^T=(\bgamma_1^T,
\ldots, \bgamma_n^T) \in\bbbr^{mn}$ denote the
stacked vectors of
the involved mask, rotation and translation parameters,
respectively, and $\bGamma_i\,{=}\,\bGamma_i(\btheta_i), i\,{=}\,1,\ldots,n$.
Moreover, $\lambda_l^i$ denotes the $l$th entry of the mask
vector $\blambda_i$, $\bx_l^i$ is the Cartesian coordinate vector of
the $l$th landmark in the $i$th point set, and
$\tilde{w}_l^i(\blambda_i)$ denotes the corresponding normalized
kriging weight. For the multiple alignment of $M_1, \ldots, M_n$ we
want to maximize (\ref{eqmultiCarbo}) with respect to the
$m(m-1)n/2+mn+\sum_i k_i$ parameters.

Define a ``normalized mean field'' of all but the $i$th point set
as
\[
\tilde{Z}_{(i)}\bigl(\bx;\blambda_{(i)},\btheta_{(i)},\bgamma_{(i)}\bigr
) =
\frac{1}{n-1} \sum_{j \neq i} \sum_{l\dvtx\lambda_l^j=1}
\tilde{w}_l^j(\blambda_j) \sigma\bigl((\bGamma_j
\bx_l^j+\bgamma_j)-\bx\bigr),
\]
where $\btheta_{(i)}^T=(\btheta_1^T, \ldots, \btheta^T_{i-1},
\btheta^T_{i+1}, \ldots, \btheta^T_{n})$,
$\bgamma_{(i)}^T=(\bgamma_1^T,
\ldots, \bgamma^T_{i-1},
\bgamma^T_{i+1}, \ldots,\bgamma^T_{n})$ and
$\blambda_{(i)}^T=(\blambda_1^T, \ldots, \blambda^T_{i-1},
\blambda^T_{i+1}, \ldots, \blambda^T_{n})$ and let
$C_{(i)}(\btheta_{i},\bgamma_i,\blambda_i;\btheta
_{(i)},\bgamma
_{(i)},\blambda_{(i)})$
denote the inner product of
$\tilde{Z}_{(i)}(\bx;\blambda_{(i)},\btheta_{(i)},\bgamma_{(i)})$
and $\tilde{Z}_{i}(\bx;\blambda_{i},\btheta_{i},\bgamma_{i})$. It
can be seen that (\ref{eqmultiCarbo}) has the property
\[
C(\btheta,\bgamma,\blambda) \propto\frac{1}{n}
\sum_{i=1}^n
C_{(i)}\bigl(\btheta_{i},\bgamma_i,\blambda_i;\btheta
_{(i)},\bgamma
_{(i)},\blambda_{(i)}\bigr).
\]
Due to this decomposition, the optimization can be carried out
stepwise by maximizing
$C_{(i)}(\btheta_{i},\bgamma_i,\blambda_i;\btheta
_{(i)},\bgamma
_{(i)},\blambda_{(i)})$
in turn. The vectors $\btheta_{(i)}$, $\bgamma_{(i)}$ and~$\blambda
_{(i)}$ are thereby kept fixed at each step.

An optimization of the overall Kernel Carbo index
$C(\btheta,\bgamma,\blambda)$ is numerically
difficult. However, we
can replace it by posterior inference within the MCMC scheme developed
for the pairwise alignment. As before, the choice of the prior
distribution for the precision parameter $\tau$ determines how much
the algorithm pushes the estimates of the other model parameters
toward the posterior mode. An iterative stochastic optimization of
the normalized fields $\tilde{Z}_{i}(\bx)$ can therefore be
formulated by employing a ``large precision version''\vadjust{\goodbreak} of the MCMC
algorithm for the pairwise alignment and then using the obtained MAP
estimates to determine a new mean field. This procedure will in
practice decrease $C(\btheta,\bgamma,\blambda)$
at every step and
can be repeated until a convergence criterion is met.

\begin{algorithm}[t]
\begin{algorithmic}[1]
\small
\vspace*{0.2cm}
\STATE choose the smallest point set as reference and superimpose the
$n-1$ remaining configurations onto it
\\[0.1cm]
\STATE define $d \leftarrow d_0,$ where $d_0 > \mathrm{tol}$ and $\mathrm{tol}$ is a
positive tolerance threshold \\[0.1cm]
\STATE calculate the multiple Carbo index $C(\btheta,\bgamma,\blambda)$
\\[0.1cm]
\WHILE{$d > \mathrm{tol} $} \vspace{0.1cm}
\FOR{$i$ in $(1\dvtx n)$} \vspace{0.1cm}
\STATE using the current parameter values for rotation, translation
and mask vectors, calculate a normalized mean field $\tilde{Z}_{(i)}(\bx
)$ omitting the $i$th configuration \\[0.1cm]
\STATE based on the dissimilarity $D_{(i)}(\btheta_{i},\bgamma
_i,\blambda_i)$, superimpose the $i$th predicted field onto $\tilde
{Z}_{(i)}(\bx)$; $\tilde{Z}_{(i)}(\bx)$ thereby takes the role of the
reference field and $\blambda_{(i)}$, $\btheta_{(i)}$ and $\bgamma
_{(i)}$ are treated as fixed \\[0.1cm]
\STATE record the MAP estimates for position and mask of the $i$th
configuration \\[0.1cm]
\ENDFOR\vspace{0.1cm}
\STATE calculate the updated Carbo index $C^*(\btheta,\bgamma,\blambda
)$ \\[0.1cm]
\STATE$d \leftarrow C^*(\btheta,\bgamma,\blambda)-C(\btheta,\bgamma
,\blambda)$
\STATE$C(\btheta,\bgamma,\blambda) \leftarrow C^*(\btheta,\bgamma
,\blambda)$
\ENDWHILE
\caption{Stochastic field GPA for unlabeled marked point sets} \label{Alg:GPA}
\end{algorithmic}
\end{algorithm}

Our field GPA algorithm is displayed as Algorithm \ref{Alg:GPA}. As the
objective of the multiple alignment of the given marked point sets
is to find the features common to all or most of the
objects, the algorithm superimposes each point set on the smallest
(in terms of the number of points) one in the data set as a first
step. Contrary to the pairwise alignment which started at a random
place in the parameter space, this initialization will be close to
the global optimum which justifies the use of the large prior mean
for the precision values. All the methods described in this paper
have been implemented in~R [\citet{R2010}], and the code can be
found in the supplementary materials [\citet{suppA}].

Note that the multiple alignment method assumes a common underlying
reference field for all point sets. However, in our steroid application
the molecules may exhibit different binding mechanisms even with the
same receptor.
In this case, several reference fields could underlie the matching
parts of the molecules. As we discuss below, we therefore consider
distinct sub-groups of molecules
(e.g., based on chemical properties or from cluster analysis)
and then look for common reference fields in various subgroups. In
other applications, a similar subgroups based approach may also be
suitable.\vadjust{\goodbreak}

\section{Simulation studies} \label{secSimulation}

\subsection{Simulation of marked point sets in two dimensions} \label
{sec2DSimulation}

\begin{figure}

\includegraphics{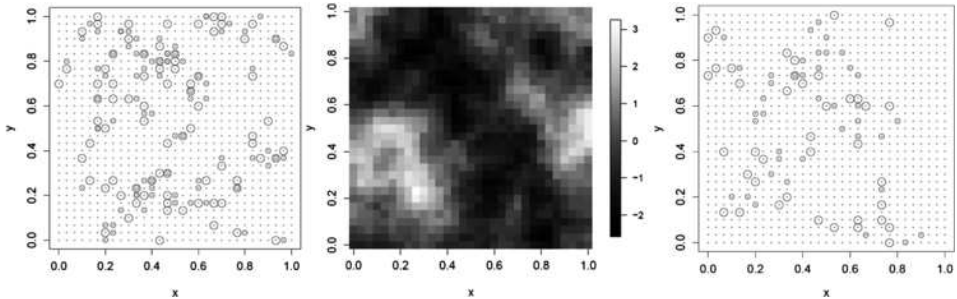}

\caption{Example of an underlying reference field
and two sampling schemes: the underlying reference field (middle) is
a realization of a zero-mean isotropic Gaussian random field with a
Mat{\'e}rn covariance function ($\nu=1$ and $\rho=0.2$). The other plots
show two sampling schemes for $B^{\rmtrue}$ (big circles)
and $A^{\rmtrue}$ (small circles): $n_{\rmB}
=n_{\rmA}=80$ and $\kappa=1$ on the left-hand side and
$n_{\rmB}=n_{\rmA}=40$ and $\kappa=4$ on the
right-hand side. The dots correspond to the 961 possible point
locations.} \label{figsimulations}
\end{figure}

We first carry out a two-dimensional simulation study to illustrate
the methodology and examine the performance of the algorithms for
different choices of parameters. We simulate marked point
sets\vspace*{1pt}
$A=\{z^{\rmA}(\bx^{\rmA}_1), \ldots,
z^{\rmA}(\bx^{\rmA}_{k_{\rmA}})\}$
and $B=\{z^{\rmB}(\bx^{\rmB}_1), \ldots,
z^{\rmB}(\bx^{\rmB}_{k_{\rmB}})\}$
which share a common underlying reference field.
As a reference field, we use a realization of a zero-mean Gaussian
random field with an isotropic Mat{\'e}rn covariance function defined
on a grid of 961 regularly spaced points
$\by_i$ within the unit square, that is, we simulate from $\tilde{\bZ}=(
\tilde{Z}(\by_1), \ldots, \tilde{Z}(\by_{961}))^T \sim\N(
\bzero, \bSigma)$,
where $\bSigma_{ij} = \sigma( \| \by_i - \by_j \|)$ is given
in~(\ref{eqmatern}).
For our simulations we use $\rho=0.2$ and $\nu=1$. Figure
\ref{figsimulations}(middle) shows a~realization $\tilde{\bz}$ of
$\tilde{\bZ}$.

%

Let $A=\{A^{\rmtrue},A^{\rmcont}\}$ and
$B=\{B^{\rmtrue},B^{\rmcont}\}$, where
``true'' denotes the part of each point set which stems from the
underlying reference field $\tilde{\bz}$
and ``cont'' denotes the
contaminated part. The term ``contaminated'' refers to the points
which do not follow the field model well and so will not be helpful in
the alignment.
Hence, the contaminated points should be masked in the matching algorithm.

We obtain\vspace*{1pt} $B^{\rmtrue}$ by randomly choosing
$k^{\rmtrue}_{\rmB}$ entries $i_1, \ldots,
i_{k^{\rmtrue}_{\rmB}}$ from $\tilde{\bz}$
and adding independent
Gaussian noise with standard deviation $\sigma_{\varepsilon}$ to
the corresponding marks.
For $B^{\rmcont}$,
$k^{\rmcont}_{\rmB}=k_{\rmB}-k^{\rmtrue}_{\rmB}$
locations on the grid are chosen at random and the corresponding
marks are random values from a~uniform distribution on $[-c,c]$. To
obtain $A^{\rmtrue}$, we introduce a nearness parameter
$\kappa\in\bbbn$ and define a set of grid points $\mathcal
{U}_{\kappa}$ as the union of neighborhoods around the points $\bx
^{\rmB}_i$ ($i=1, \ldots, k^{\rmtrue}_{\rmB}$), where each
neighborhood contains the vertically,
horizontally and diagonally adjacent grid points in a~$(2\kappa
+1)\times(2\kappa+1)$-box around the corresponding $\bx^{\rmB}_i$.
The parameter $\kappa$ therefore measures the
nearness between points in terms of grid point locations rather than
Euclidean distance which is further demonstrated in Figure~\ref
{figsimulations}. The point locations $\bx^{\rmA}_j$
($i=1, \ldots, k^{\rmtrue}_{\rmA}$) which
belong to the matching part of $A$ are then chosen at random from
$\mathcal{U}_{\kappa}$ and $A^{\rmtrue}$ is obtained by
adding independent Gaussian noise with
standard deviation $\sigma_\varepsilon$ to the corresponding marks
$\tilde{z}(\bx^{\rmA}_j)$ ($i=1, \ldots, k^{\rmtrue}_{\rmA}$).
Finally, the
$k^{\rmcont}_{\rmA}=k_{\rmA}-k^{\rmtrue}_{\rmA}$ points
in~$A^{\rmcont}$ are obtained in the same way as the
contamination points in $B$.

Note that this simulation scheme does not create pairwise
correspondences between points in $A^{\rmtrue}$ and
$B^{\rmtrue}$.
Although we have used a nearness criterion in our simulation method,
we have not estimated point correspondences in the course of the MCMC algorithm.


For our simulation study we consider three realizations of
$\tilde{\bZ}$, and for each of these realizations we define 12
different pairs of marked point sets by varying the parameters
$k^{\rmtrue}=k^{\rmtrue}_{\rmA}=k^{\rmtrue}_{\rmB}\in\{40,80\}$,
$k^{\rmcont}=k^{\rmcont}_{\rmA}=k^{\rmcont}_{\rmB}\in\{
0.05k^{\rmtrue},0.1k^{\rmtrue},0.15k^{\rmtrue}\}$ and
$\kappa\in\{1,4\}$. Moreover, we choose \mbox{$c=7$} and $\sigma_{\varepsilon
}=\sqrt{0.02}$. Generated as above, the 36 pairs $A$ and
$B$ are recorded in the optimal relative position, and the optimal mask
vectors are
$\blambda_{\rmA}^T=(\bone_{k^{\rmtrue}_{\rmA}}^T,\bzero
_{k^{\rmcont}_{\rmA}}^T)$
and
$\blambda_{\rmB}^T=(\bone_{k^{\rmtrue}_{\rmB}}^T,\bzero
_{k^{\rmcont}_{\rmB}}^T)$.

\subsection{Hyperparameter settings}

For each pairwise superposition 50,000\break MCMC iterations are carried
out, and each iteration contains five blocks updating the rotation
parameter (proposal standard deviation: $0.75^{\circ}$), the
translation vector (proposal standard deviation:
0.01), the precision parameter and the two mask vectors,
respectively. The Kernel Carbo similarity calculations are based on the
exponential kernel, that is, (\ref{eqmatern}) with $\nu=0.5$ (whereas
$\nu=1$ was used for simulating the data).
Initially we use $\rho=0.6$, but, within the first
1,000 iterations, this value is dynamically reduced to $\rho=0.2$.
This initial phase allows the algorithm to home in on a good
alignment even if the two points sets\vspace*{1pt} are far away from their
optimal relative position. Moreover, we use
$\beta=0.05$ and $\alpha=200$, and these values ensure a~desirable
interaction between the obtained dissimilarity values and the proposed
precision values at each iteration. We include $\zeta$ as a variable
parameter in our simulation
study and consider $\zeta\in\{10,50,90\}$, and we fix $\zeta_I =
1$.\looseness=-1



To overcome the difficulty of getting trapped in local modes, we
propose a big change
of the rigid body parameters by using increased values for the standard
deviations
of the random walk proposals every 125 iterations. Moreover, we restart
the algorithm if the
Carbo distance exceeds 0.3 after 7,500 iterations.

\subsection{Results}

For each of the 108 (36 pairs of point sets $\times$ 3 values of
$\zeta$) considered MCMC\vadjust{\goodbreak} runs, the starting position of the movable
point set $B$ is obtained by rotating and translating it from its
original (simulated) position using $\bGamma(\theta_0)$ and $\bgamma
_0$ where ${\theta_0}$ and ${\gamma_0}_i$ $(i=1, 2)$
are uniformly distributed on $[-20^{\circ},20^{\circ}]^3$ and
$[-0.1,0.1]$, respectively. Moreover, both mask vectors are initiated
using $\lambda_i^{\rmM} \sim\operatorname{Bernoulli}(0.5)$
($i=1, \ldots, k_{\rmM}$, $M \in\{A,B\}$). The
performance of each run is then assessed in terms of the root mean
square deviation (RMSD) between the original position of $B$ and its
MAP position. 

%
\begin{figure}

\includegraphics{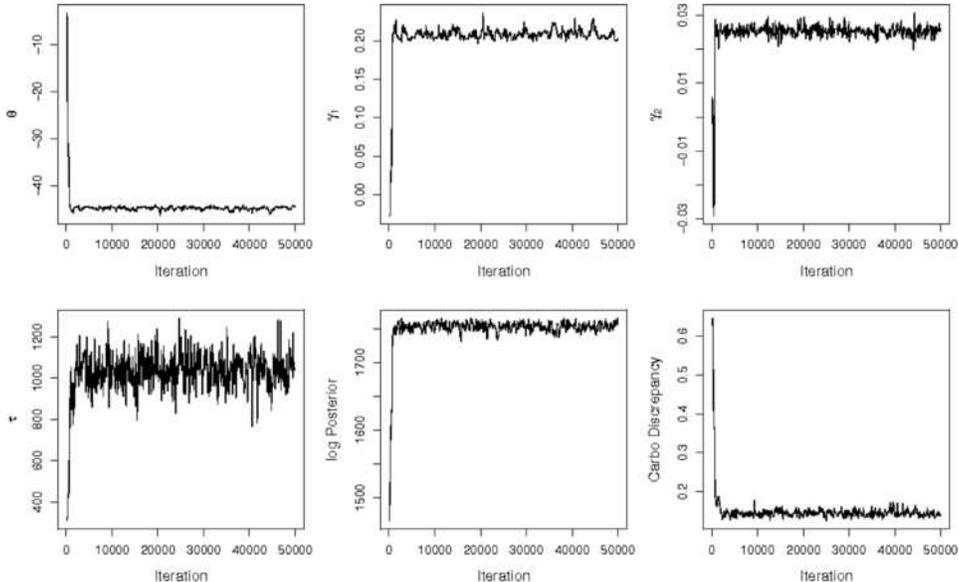}

\caption{Top row: trace plot of the rigid-body
parameters (in terms of the initial relative position of the two
points sets under consideration). Bottom row: trace plots of the
precision parameter, the log-posterior (up to a constant) and the
Kriged Carbo distance. In all plots, every 100th simulated value is
displayed.} \label{figmonsum}
\end{figure}

\begin{figure}

\includegraphics{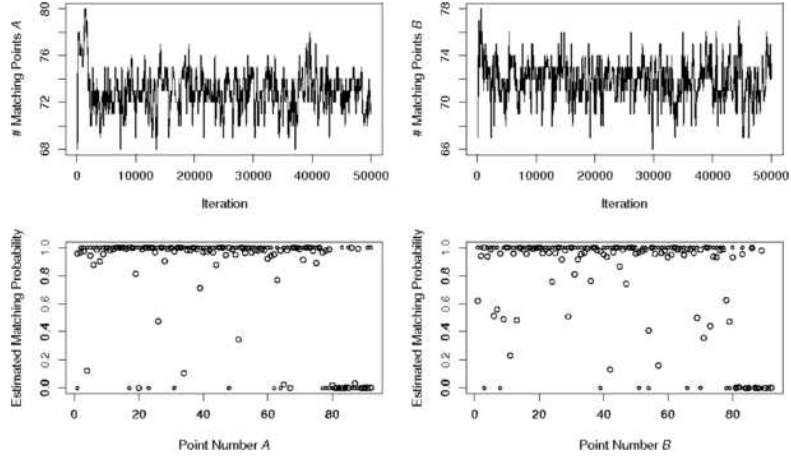}

\caption{Top row: trace plots of the number of
points involved in the kriging procedure. Bottom row: two possible
point estimates for the mask vectors of $A$ (left) and $B$ (right).
The big circles show the mean values of the $(0,1)$-entries for the
masks vectors (which can be interpreted as the estimated probability
for the corresponding landmark to belong to the common reference
field), and the small circles display the observed mask vectors at
the MAP iteration. The total number of points in $A$ and $B$ is 92,
and the last 12 points in each set are contamination points.}
\label{figmaskssim}
\end{figure}

\begin{figure}[b]

\includegraphics{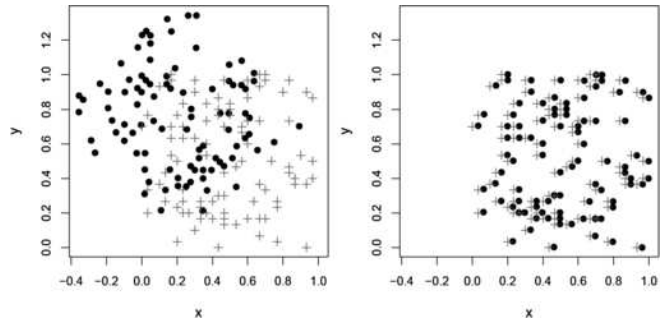}

\caption{Successful alignment: the circles on the left-hand side
show the initial position of point set $B$, and the circles on the
right-hand side show the position
of $B$ at the MAP iteration. The optimal position is displayed by the
crosses on both sides. The algorithm is able to reduce the
RMSD to the optimal position from 0.479 (left) to 0.032 (right).}
\label{figresultsim}
\end{figure}

Figures \ref{figmonsum}--\ref{figresultsim} show the typical output of
a successful run.
Figures \ref{figmonsum} and \ref{figmaskssim} indicate that the
algorithm converges quickly, and from Figure \ref{figmonsum} it can be
seen that there is an
interplay between the precision parameter $\tau$ and the Kernel Carbo
distance: until a good alignment is obtained, small distances lead to
larger precision values which in turn yield small distances. Figure
\ref{figmaskssim} shows the trace plots for the number of points which
are involved in the field calculation and are hence considered to
belong to $A^{\rmtrue}$ and $B^{\rmtrue}$,
respectively, and
a (post burn-in) summary of the two mask vectors is displayed in the
bottom row of Figure \ref{figmaskssim}. In this particular example,
the optimal mask vectors are $\blambda_{\rmM}^T=(\bone
_{80}^T,\bzero_{12}^T)$ ($M=A,B$), and the algorithm is able to
reconstruct the mask vector very well. Figure \ref{figresultsim} shows
that the MAP position of the movable point set is very similar to the
original one.

%

%

We consider an alignment to be successful if $\rmsd\le0.1$. Table
\ref{TAB1} shows the percentages
of success for various parameter settings in the row ``setting~1,''
with an overall success rate of 76\% out of the 108 MCMC runs.
As expected, the largest number of true points in combination with the
fewest number of
contamination points gives the highest success rate
(89\%), whereas the smallest number of true points in combination with
the highest number of contamination
points gives the lowest success rate ($44\%$). In combination with
these extreme cases, the impact of the nearness parameter is most
striking with $22\%$ success for ($k^{\rmtrue}=40,
k^{\rmcont}=6$, $\kappa=4)$ and $100\%$ for ($k^{\rmtrue}=80,
k^{\rmcont}=4, \kappa=1)$. Overall,
the impact of $\kappa$ can be summarized as $85\%$ success for $\kappa
= 1$ and $67\%$ success for $\kappa=4$. Interestingly, the success
rate increases with $\zeta$.

%
\begin{table}
\tabcolsep=5pt
\caption{The percentages of MCMC runs which are regarded
as a success (i.e., $\rmsd< 0.1$) for different parameter settings.
The column ``all'' shows the percentage of successes
out of all 108 simulations for the corresponding setting}\label{TAB1}
\begin{tabular*}{\tablewidth}{@{\extracolsep{\fill}}lcccccccc@{}}
\hline
& & & & & $\bolds{k^{\rmtrue} = 80}$ & $\bolds{k^{\rmtrue} = 40}$
& &\\
& \textbf{All} & $\bolds{\zeta=10}$ & $\bolds{\zeta= 50}$ & $\bolds
{\zeta= 90}$
& $\bolds{k^{\rmcont} = 4}$ & $\bolds{k^{\rmcont} = 6}$ & $\bolds{\kappa
=1}$ &
$\bolds{\kappa=4}$\\
\hline
Setting 1 & 76 & 61 & 81 & 86 & 89 & 44 & 85& 67 \\
Setting 2 & 48 & 33 & 47 & 61 & 83 & 17 & 52 & 44\\
\hline
\end{tabular*}
\end{table}

The above results indicate that a satisfactory alignment can be
obtained if the number of noncontamination points is large enough
to represent the main features of the underlying reference field and
large relative to the number of contamination points. Moreover, especially
when the number of points is small and the sampling of the reference field
is sparse, it is important that the noncontamination points in $A$ and $B$
represent the same features of the reference field (which is not always
the case if $k^{\rmtrue}=40$ and $\kappa=4$). These
trends can
be emphasized by rerunning the experiments
using $\theta\sim U_{[-60^{\circ},60^{\circ}]}$ and
$\gamma_i \sim U_{[-0.3,0.3]}$ $(i=1,2)$ to obtain the starting
position of $B$.
For this more challenging setting (``setting 2'') the results are also
provided in Table~\ref{TAB1} with similar effects
but lower success rates (48\% overall).

In both settings, the performance of our alignment procedure can
be improved if there are some points in $A$ and $B$ which can be
identified as noncontamination points \textit{ab initio}.
For our examples, identifying some relevant points (on average 12
per point set) improves the overall success rate to 93\%
in the first setting and to 79\% in the more challenging second setting.
In many applications it may be possible to identify some relevant
points so that the possibility of incorporating this knowledge is a
valuable tool for improving the alignment. 

Finally, we rerun the above experiments with different values for the range
parameter $\rho$. For example, with $\rho=0.3$, overall success rates of
77\% in the first and 48\% in the second setting
are achieved, and for $\rho=0.1$, the corresponding success rates are
77\% and 52\%.
These results demonstrate that choosing the exact covariance function
for the spatial interpolation is not crucial for the performance of the
algorithm,
although performance does deteriorate for much larger $\rho$.
For example, a leave-one-out type method for identifying the contamination
points combined with a pooled version of an
experimental semivariogram [e.g., \citet{Wackernagel2003}, page
47] can be
applied to estimate an approximate
covariance function which has yielded satisfactory results in some
further experiments.

\subsection{Three-dimensional example}\label{sec4.4}

We now consider a small three-dimen\-sional simulation study which
mimics the
molecule alignment problem of Section~\ref{secApplication}. As a
starting point we take the positions of the first $25$
atoms of the first molecule in the steroid
data set and generate the atom positions of a second ``molecule'' using a
small perturbation (independent zero mean normal
with standard deviation $0.01$). Then a zero mean isotropic Gaussian
random field with
Mat{\'e}rn
covariance function $(\nu=0.5,\rho=5)$ is simulated at the combined
set of the 50 points. To introduce contamination points, the last five
points in each configuration have their coordinates and marks perturbed
by independent $N(0,3^2)$ errors. Finally, both molecules are centered
and the molecules are
uniformly rotated.

For various
choices of the hyperparameters $\beta$ and $\zeta$ we run $100$ Monte
Carlo simulations of the Bayesian alignment procedure. Each time the
two marked point sets and their starting relative position are
generated as above. The parameters $\nu=0.5$ and $\alpha=31$ are kept
fixed and the range parameter is dynamically reduced from $\rho=20$ to
$\rho=5$ during the matching procedure.
Each simulation is restarted if the Kernel Carbo distance is greater
than $0.1$ after $1\mbox{,}000$ iterations
(up to a maximum of $30$ restarts). When the algorithm reaches $2\mbox{,}000$
iterations
the final position and the MAP position of the movable molecule $B$ are
recorded.

In this situation the success of the algorithm can be
measured in terms of the first 20 atoms of $B$ by taking the
corresponding $\rmsd$
between its MAP and its true position. The results of the simulation study
are given in Table \ref{TAB2}.
As expected, the number of unmasked points in $B$ increases with $\zeta
$. Interestingly, this consistently also leads to improved $\rmsd$
values---even in situations where a large value of $\zeta$ forces the
algorithm to include more than the desired 20 points.
In terms of the obtained Carbo distance, the impact of $\beta$ exceeds
that of $\zeta$. This is also expected, as the mean of the full
conditional distribution of the precision parameter $\tau$ (cf.
Section \ref{subpostsampling}) decreases with $\beta$ which in turn
means that the
algorithm is more prone to accept updates with larger Carbo distances.

Overall, this simulation study highlights that the Bayesian method
works well in this controlled situation.

%
\begin{table}
\caption{Summary statistics from the posterior distribution in the
simulation study. Columns 2--6 show the mean (and standard deviation)
over 100 Monte Carlo\vspace*{1pt} simulations of the final number of
unmasked points in molecule $A$ ($\sum\lambda^{\rmA}_{i}$); the final
number of unmasked points in molecule $B$ ($\sum\lambda^{\rmB}_{j}$);
the root mean square error ($\rmsd$); the number of new starts needed
for the algorithm to be successful; and the Kriged Carbo distance at
the final iteration. The last column shows the number of times out of
$100$ simulations that the algorithm failed, that is, the Kernel Carbo
distance was greater than $0.1$ after $1\mbox{,}000$ iterations
for each of the maximum number of 30 restarts}\label{TAB2}
\begin{tabular*}{\tablewidth}{@{\extracolsep{\fill
}}ld{2.3}d{2.3}d{2.5}d{2.3}d{2.5}c@{}}
\hline
\multicolumn{1}{@{}l}{$\bolds{(\beta,\zeta)}$} &
\multicolumn{1}{c}{$\bolds{\sum\lambda^{\rmA}_{i}}$\hspace*{1.5pt}}
& \multicolumn{1}{c}{$\bolds{\sum\lambda
^{\rmB}_{j}}$\hspace*{1.5pt}}
& \multicolumn{1}{c}{$\bolds{\rmsd}$} & \multicolumn{1}{c}{\textbf{Starts}}
& \multicolumn{1}{c}{\textbf{Carbo}} & \multicolumn{1}{c@{}}{\textbf
{Failures}}\\
\hline
$(0.0004,10)$ & 18.41 & 17.42 & 0.1523 & 0.88 & 0.0204 & 4\\
& (2.39)&(2.13) & (0.5207) & (2.48) & (0.0226) & \\
$(0.0004,50)$ & 21.16 & 20.00 & 0.0959 & 1.56 & 0.0178 & 1\\
& (1.45)& (0.97)& (0.6521) & (3.96) & (0.0195) & \\
$(0.0004,70)$ &21.66 & 20.43 & 0.0626 & 1.12 & 0.0263 & 0\\
& (1.53) & (0.97) & (1.1200) & (2.80) & (0.0208) & \\
[4pt]
$(0.004,10)$ & 18.6 & 17.83 & 0.2193 & 0.67 & 0.0268 & 0\\
& (2.51) & (1.90) &(0.5492)&(1.21)& (0.026) & \\
$(0.004,50)$ & 21.52 & 20.27 & 0.1018 & 1.55 & 0.0284 & 1 \\
& (1.46 & (1.08) & (0.4352) & (3.80) & (0.0560) & \\
$(0.004,70)$ & 21.58 & 20.32 & 0.0605 & 1.19 & 0.0280 & 0\\
& (1.42) & (1.17) & (0.0734) & (3.34) & (0.0244) &\\
[4pt]
$(0.04,10)$ & 20.90 & 19.47 & 0.1306 & 0.95 & 0.0342 & 0\\
& (1.75) & (1.60) & (0.5884) & (1.57) & (0.0187) & \\
$(0.04,50)$ & 23.03 & 20.94 & 0.0739 & 1.59 & 0.0485 & 1\\
& (1.35) & (1.19) & (0.2748) & (3.75) & (0.0544) &\\
$(0.04,70)$ & 23.15 & 20.92 & 0.0513 & 2.26 & 0.0472 & 0 \\
& (1.37) & (1.04) & (0.0629) & (4.41) & (0.0258) & \\
\hline
\end{tabular*}
\vspace*{3pt}
\end{table}

\section{Application to steroid molecules} \label{secApplication}
The concept of molecular similarity is of great importance because
similar molecules can be expected to exhibit a similar biochemical
activity and hence drug potency. The data for the 31
steroids considered by \citet{Drydenetal07} are given in the form
of a set of unlabeled, marked points
where the coordinates of the points correspond to the atom coordinates
of each molecule, and the marks are the partial charge values and the van
der Waals radii. The data set can be found in the supplementary
materials [\citet{suppB}].
The Kernel Carbo index developed in this paper can therefore
directly be utilized to assess the similarity between the steroids.
Also, in particular, the assumption of a common underlying
reference field seems suitable for this application because all
molecules bind to the same receptor protein. The underlying
reference field can therefore be interpreted as a negative imprint
of the binding pocket of the receptor. The MCMC scheme described in
Section \ref{secparwise} then determines the parts of each molecule
which correspond to the reference field and aligns the molecules
based on the similar parts only so that the resulting relative
position should reproduce the relative binding positions of the
steroids.

In order to investigate the
possibility of multiple binding modes (and hence reference fields), we
shall also consider an analysis of
subgroups of the data. In particular, we consider the three activity classes.

\subsection{Pairwise alignment} \label{subresultpairs}

In our application we use the Gaussian kernel (\ref{Gaussian-kernel}) for
the spatial interpolation of
both the partial charge values and the van der Waals radii. The
range parameter $\rho$ for the electrostatic field is thereby
estimated by visual inspection of a pooled empirical semivariogram
function ($\rho_Q=6.35$), and the practical range of the steric
(shape) field is\vspace*{1pt}
taken to be the largest van der Waals radius in the data set
($\rho_S = 1.7/\sqrt{3} = 0.9815$).

In our simulation studies we dynamically reduced the
range parameter to help the algorithm home in on a good solution.
Here, we apply a~different concept using a weighted average of the two
univariate partial Carbo indices and choosing the weights
dynamically as $w_Q = \frac{N_I-i}{N_I}$ and $w_S
= \frac{i}{N_I}, i=1, \ldots, N_I$, during\vspace*{2pt} an initial phase
of $N_I=1\mbox{,}500$ iterations. This directly mimics real-life molecular
recognition where the long-range electrostatic attraction governs
the initial approach of the molecules, whereas the short-range
repulsive steric forces gradually take over and become the chief
manipulator for the binding affinity [e.g., \citet{Richards1993}].

We use
$\alpha=31$ and $\beta=0.04$, and the value for the penalty
parameter is chosen as $\zeta=3$. As standard deviations of the
proposal distributions we use $\eta_1=3.25^{\circ}$ for the rotation
parameters and $\eta_2=0.25 \mbox{ \AA}$ for the translation
parameters, and these values ensure acceptance rates between 20\%
and 40\%. The standard deviation for the rotation parameters is
thereby in line with previously described proposal distributions for
rotation parameters in the molecular context
[e.g., \citet{Greemard06}]. We define the initial relative
position of two molecules by first aligning both molecules along
their principal axes. We then translate and rotate the movable
molecule using $\bgamma_0$ and $\bGamma(\btheta
_0)$, where\vspace*{1pt}
${\gamma_0}_i$ $(i=1, 2, 3)$ and ${\theta_0}_i$ $(i=1, 2, 3)$ are
uniformly distributed on $[-5\mbox{ \AA},5\mbox{ \AA}]$ and
$[-90^{\circ},90^{\circ}]$, respectively.

%
\begin{figure}

\includegraphics{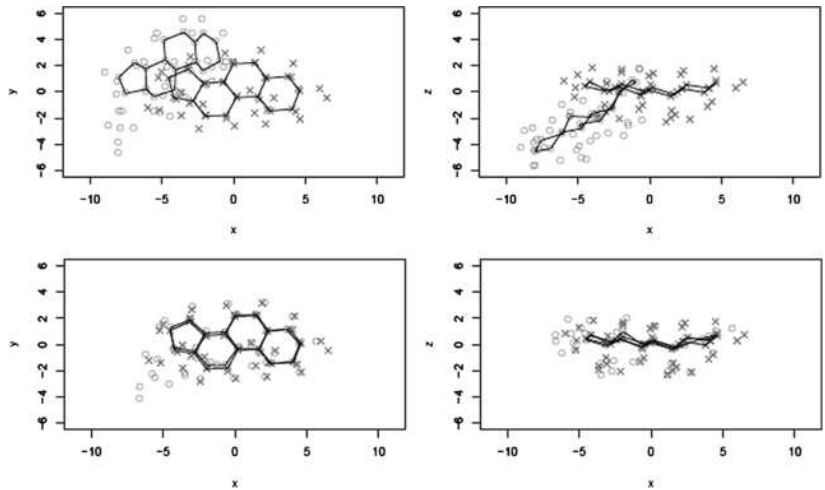}

\caption{Two orthographic projections ($x$--$y$ and $x$--$z$ Cartesian
planes) of the atoms in
the starting position (top row) and the MAP position for the alignment
(bottom row)
of steroid molecules aldosterone and androstanediol. The carbon rings
are indicated
by lines for each molecule.
The unit of all axes is \AA ngstr\"om (\AA).} \label{figpositions}
\end{figure}

As our MCMC algorithm is asymmetric in the sense that the relative
position of the molecules is changed by moving only one molecule
whereas the position of the other one is fixed, we consider all $31
\cdot30=930$ pairwise superpositions. In the majority of cases, the
algorithm converges quickly and the trace plots show a similar
behavior as the ones in Figures \ref{figmonsum} and~\ref{figmaskssim}. However, like in the simulation studies, the
algorithm can sometimes get trapped in a local mode (which mostly
corresponds to an alignment along the wrong principal axes in this
application) so that a restart is
necessary.
Figure~\ref{figpositions} shows an example result where the steroid aldosterone
has successfully been superimposed onto androstanediol.

The specific choice of Gaussian kernel is not
crucial to the success of the
algorithm for the steroids data.
Similar alignments of the steroids have been obtained using the Mat\'ern
kernel with $\nu= 0.5, \nu=1, \nu= 2$ for many examples, but it is
important that
the range parameter is well chosen. We found that with $\rho_S \approx1$
the method worked well for any choice of covariance kernel
we used, but if $\rho_S$ is too large (e.g., $\rho_S \approx3$),
then
the alignment is cruder and the algorithm is more prone to failure for
any choice of covariance function.\looseness=-1\vadjust{\goodbreak}

\subsection{Prior sensitivity}
To investigate the sensitivity of the
analysis to the prior distributions, we again consider the alignment
of the two molecules aldosterone and androstanediol. Table
\ref{tabimpact} shows how different values of the penalty parameter
%
%
\begin{table}[b]
\caption{The impact of the penalty parameter (first four rows) and
$\alpha$ (last four rows) on the marginal posterior distribution of
the parameters of interest. The credibility intervals are based on
every 20th value of the parameters recorded after the burn-in
period} \label{tabimpact}
\begin{tabular*}{\tablewidth}{@{\extracolsep{\fill}}lccc@{}}
\hline
$\bolds{\zeta}$ & \textbf{95\% CI for} $\bolds{\tau}$ & \textbf{95\%
CI for} $\bolds{\sum_j \lambda^{\rmA}_j}$ & \textbf{95\%
CI for} $\bolds{\sum_j \lambda_j^{\rmB}}$\\
\hline
\hphantom{0}2 &$(226.62, 543.78)$ & $(34, 46)$ &
$(34, 45)$\\
\hphantom{0}3 &$(230.93, 543.30)$ & $(37, 49)$ &
$(38, 48)$\\
\hphantom{0}4 &$(250.69, 562.65)$ & $(40, 51)$ &
$(40, 49)$\\
\hphantom{0}5 &$(244.67, 548.41)$ & $(41, 51)$ &
$(42, 51)$\\
\hline
$\bolds{\alpha}$ & \textbf{95\% CI for} $\bolds{\tau}$ & \textbf{95\%
CI for} $\bolds{\sum_j \lambda^{\rmA}_j}$ & \textbf{95\%
CI for} $\bolds{\sum_j \lambda_j^{\rmB}}$\\
\hline
21 &$(102.53, 315.95)$ & $(36, 48)$ &
$(37, 48)$\\
31 &$(221.14, 515.13)$ & $(38, 49)$ &
$(38, 49)$\\
41 &$(344.68, 770.30)$ & $(38, 48)$ &
$(39, 49)$\\
51 &$(432.36, 1010.77)$ & $(35, 48)$ &
$(37, 50)$\\
\hline
\end{tabular*}
\vspace*{-3pt}
\end{table}
$\zeta$ affect the empirical (post burn-in) 95\% credibility
intervals for the number of included atoms for both molecules based
on 10,000 iterations. As
expected, the total number of included atoms increases with~$\zeta$.
As the two molecules in the example run are structurally very
similar, they can be aligned more closely if more atoms are included
so that the credibility interval for the precision parameter is shifted
toward higher values as~$\zeta$ increases. After a certain
threshold, however, even larger values for the penalty parameter
force the algorithm to include more atoms in the similarity
calculations than desired and the precision decreases. Moreover,
Table~\ref{tabimpact} shows that---in terms of the number of
included atoms---the algorithm is robust against changes of
$\alpha$. Also, as the posterior mean and variance of the precision
parameter directly depend on $\alpha$, the credibility intervals for
$\tau$ become wider and get shifted toward higher values as
$\alpha$ increases.

\subsection{Chemical relevance}
The pairwise distances
which result from the 930 superpositions can be regarded as
chemically meaningful if they reflect the membership of the steroid
molecules to the three activity classes, that is, if steroids within an
activity class can be aligned more closely than those from different
activity classes. In terms of our assumption about a~common
underlying reference field, such a result would indicate that there
are actually three different reference fields which exhibit
different small scale variations and hence different abilities to
fit in to the protein binding pocket.

We assess the chemical relevance of our results by performing two
cluster analyses using Ward's (\citeyear{Ward1963}) method. To
account for the asymmetry in our alignment method, the applied
pairwise dissimilarity measures for two molecules $A$ and $B$ are
thereby based on both the MCMC run which superimposes $A$ on $B$ and
the MCMC run which superimposes $B$ on $A$. In particular, we use
$D_{\rmmean}(A,B)=\sqrt{\hat{D}^{\rmmean}_{A
\rightarrow B} \hat{D}^{\rmmean}_{B \rightarrow A}}$ and
$D_{\rmMAP}(A,B)=\sqrt{\hat{D}^{\rmMAP}_{A
\rightarrow B} \hat{D}^{\rmMAP}_{B \rightarrow A}}$,
where the
arrow denotes the direction of the superposition, and
``mean'' and ``MAP'' indicate which type of (post burn-in) point
estimate for the parameters is inserted into the Carbo dissimilarity measure~(\ref{eqplugdist}).

%
\begin{figure}

\includegraphics{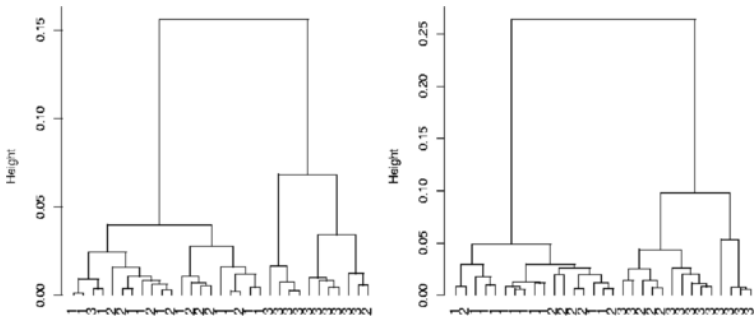}

\caption{Cluster analysis using Ward's method: the
left-hand side dendrogram is based on $D_{\rmmean}(\cdot)$,
and the dendrogram on the right-hand side is calculated using
$D_{\rmMAP}(\cdot)$. The labels correspond to the activity
classes of the steroids (1${}={}$high, 2${}={}$intermediate,
3${}={}$low).} \label{figdendrogram}
\end{figure}

Figure \ref{figdendrogram} shows the dendrograms resulting from the
cluster analyses.
It is notable that both distance measures lead to a very
good separation of high and low activity steroids. In particular,
the cluster analysis based on $D_{\rmMAP}(\cdot)$ is at the
highest level able to separate these two activity classes
completely. Overall, our distance can separate the activity classes
as well as the distance which \citet{Drydenetal07} found
to have the highest separation power, and it clearly outperforms the
other distances defined in their paper.

The dendrograms indicate that it is plausible to assume that
there are at least two different reference fields underlying the
steric properties of the steroids. It is therefore of interest
to determine these fields and examine where differences occur, as
they could give rise to the
different binding activities. In the following we will do so in a
two-step procedure where our field GPA\vadjust{\goodbreak} approach is first applied to
all 31 steroids to obtain the overall optimal relative position of the
molecules and then to the subgroups as defined by the activity classes
which will provide the appropriate masks.

\subsection{Overall multiple alignment}

When carrying out the overall optimal alignment of all 31 steroid
molecules, the pairwise superpositions
in step 1 of Algorithm~\ref{Alg:GPA} are performed as described before but
with $\zeta=2$ to incorporate the knowledge that the reference
molecule in all superpositions has a small number of atoms. The
superpositions on the mean fields (step 7) are obtained using only
the dissimilarities of the steric fields. As the initial molecular
fields obtained in step 1 are good approximations of the fields
which minimize the multiple Kernel Carbo index, we use $\alpha=600$ and
$\beta=0.0001$ to ensure that the full conditional distribution of the
precision parameter has a large mean value at each iteration, and we
reduce the standard deviations of the proposal distributions for the
rigid body parameters to $\eta_1=0.75$ \mbox{\AA} and $\eta_2 =
0.03^{\circ}$. Moreover, we set the number of iterations for each
MCMC run in step 7 to 500, and the tolerance value to $\mathrm{tol}=0.0001$.
The algorithm is therefore used as a stochastic optimizer.

%
\begin{figure}

\includegraphics{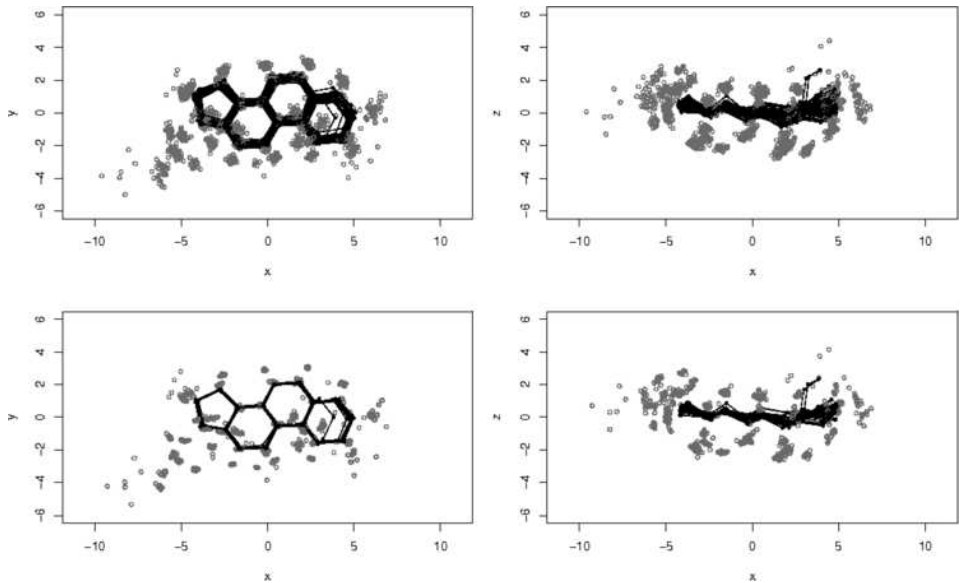}

\caption{Top row: orthographic projections of the
relative position of the 31 steroid molecules that results from step 1
of Algorithm \protect\ref{Alg:GPA}. Bottom row: orthographic
projections of the
final relative position. The random starting positions of the molecules
are not displayed.} \label{figGPAoverlays}
\end{figure}

The algorithm converges after the fourth field GPA iteration. Figure
\ref{figGPAoverlays} shows orthographic views of the resulting
overlays, that is, projections of the three-dimensional
data into the $x$--$y$ and $x$--$z$ Cartesian planes. The superposition
after step 1 of the field GPA algorithm
is displayed in the top row, and the bottom row shows the final
overlay. For clarity, the random starting positions of the
steroids are not displayed in this picture.

\subsection{Alignment within activity class subgroups}
We now carry out the field GPA algorithm in subgroups of the data to
allow for the
possibility of different underlying multiple fields. Specifically,
we consider the three activity classes of high, medium and low binding
affinity to the receptor.
The estimated\vadjust{\goodbreak} mask vectors from each underlying field are then
used together with the relative position of all molecules obtained in the
overall field GPA to calculate mean fields for each group.

\begin{figure}

\includegraphics{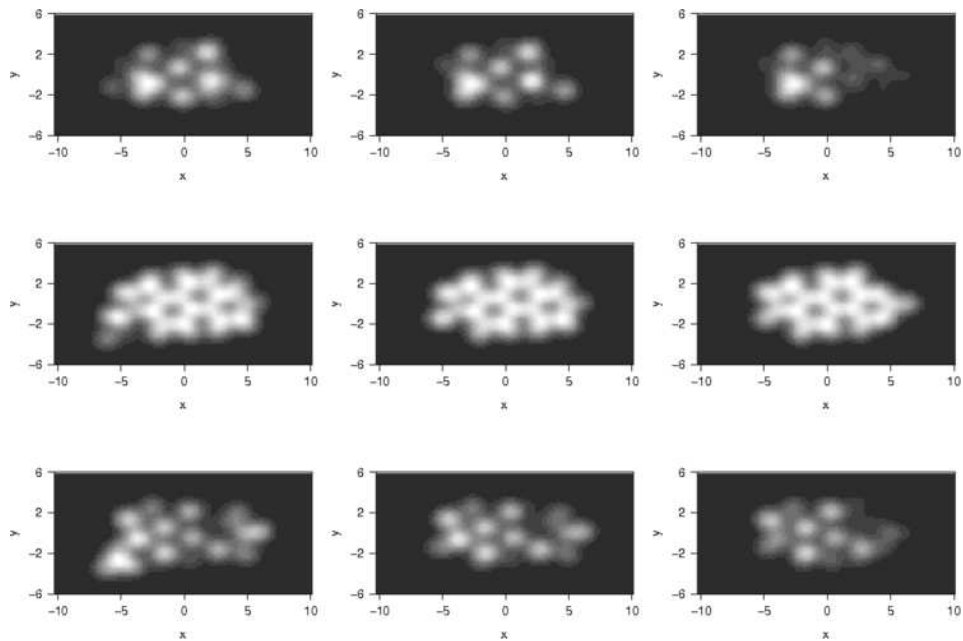}

\caption{Cross sections of the mean steric fields
of the three activity groups (left column: high activity, middle column:
medium activity, right column: low activity). The different rows
display cross sections at $z=-1.5$ (top row), $z=0$ (medium row)
and $z=1.5$ (bottom row). Light points correspond to locations with
large value of the
displayed field, whereas dark values show points with values close to
zero.} \label{figGPAgroupfields}
\end{figure}

Figure \ref{figGPAgroupfields} displays
different cross sections of the mean field for each activity class.
Light points thereby correspond to locations where the
displayed steric field takes a large value, whereas dark points show
field values close to zero.

%

%
\begin{figure}[b]

\includegraphics{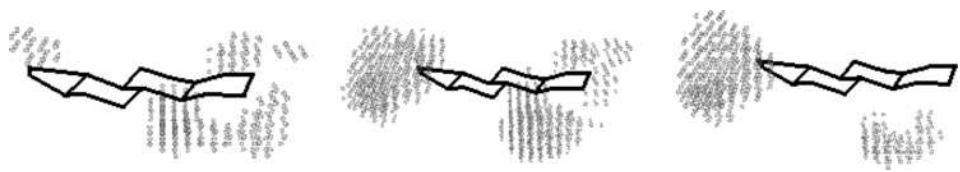}

\caption{Thresholded $t$-fields resulting from
pairwise comparisons of the steric mean fields of the three activity
classes. Left-hand side: low vs. medium activity class, middle: low
vs. high activity class, right-hand side: medium vs. high activity
class. The shaded areas display regions where the $t$-field takes
absolute values of larger than eight.} \label{figsigregions}
\end{figure}

As expected, the observed differences are most pronounced between the
mean fields of the high and the low activity group. To assess the
differences for each pair ($C_a,C_b$) of activity
classes ($a,b=1,2,3; a\neq b$) numerically,
we consider a (two sample) $t$-field of the form
%
%
\begin{equation} \label{eqt}
t_{ab}(\bx) = \frac{\bar{Z}_a(\bx) - \bar{Z}_b(\bx)}{s^*_{\rmpool
}(\bx)\sqrt{{1/n_a}+{1/n_b}}},\qquad
\bx\in\bbbr^3,
\end{equation}
where $n_a$ and $n_b$ denote the number of molecules in activity
class $C_a$ and~$C_b$, respectively, $\bar{Z}_a(\bx)$ and
$\bar{Z}_b(\bx)$ denote the corresponding mean fields, and
$s^{*2}_{\rmpool}(\bx)=s^2_{\rmpool}(\bx)+d$
is the pooled variance (with $d=0.001$ a small offset to avoid
spuriously large values in regions far away from the center).
For each pairwise comparison we define
a three-dimensional grid $G$ and calculate a~$t$-value of the form
(\ref{eqt}) at a large number of points ($142\mbox{,}598$ here).
Here we use (\ref{eqt}) as an exploratory tool to see where the
most pronounced differences occur. Figure \ref{figsigregions}
shows the regions in which the (absolute) $t$-field for each
comparison exceeds a threshold of $8$. A formal test which takes into
account the multiple comparison problem and the spatial smoothness of
the $t$-field could be
applied using a threshold based on the excursion sets of Gaussian
fields [e.g., \citet{Worsley1994}, \citet
{TaylorWorsley2008}], which has
been extensively used in
fMRI studies.

From both Figures \ref{figGPAgroupfields} and
\ref{figsigregions} it can be seen that the main feature which
distinguishes the high activity class from the other two classes is
that the very active molecules commonly extend to the left of the ring
structure much more than the other molecules,
where by ring structure we mean the carbon rings as shown in Figure
\ref{figpositions}. From the original data we can get the additional
information that the associated atoms are oxygen and carbon atoms.
Another interesting difference is located at the top right-hand side
of the molecules where the low activity class differs from the other
two classes in the location of oxygen atoms.
These findings are in line with
Figure 9 in \citet{Drydenetal07} and support
the conjecture that the steric properties of the steroid molecules
have a discriminating effect with respect to the binding affinity
toward the CBG receptor.

\section{Discussion} \label{secdiscussion}
Our methodology for aligning and comparing unlabeled marked point
sets is based on spatial interpolation of the given marks and hence
on a continuous representation of shape. The major advantages of our
approach are that point correspondences do not
need to be estimated and that the incorporated
mask vectors automatically determine the
similar regions of the considered point sets while ignoring the
rest, which helps to reduce the level of noise in the alignment procedure.

Our approach is related to a number of previously proposed methods.
For example, it provides a probabilistic framework and
generalization of the SEAL algorithm [\citet{KearsleySmith1990}]
which is well established in the field of rational drug design and
essentially uses the $L_2$-Carbo index together with a Gaussian
covariance function.
Our multiple alignment approach is related to the Bayesian model
proposed by \citet{Drydenetal07} which uses a similar
concept but formulated only in terms of the point locations. Contrary to
that, a hidden point configuration in the fully model-based
Bayesian approach by \citet{RuffieuxGreen2009} is integrated out
and the multiple alignment of $n$ point sets involves all $2^n-n-1$
possible types of matches. The fact that our field-based approach
naturally incorporates the additional information given by the marks is
an additional difference to the previous approaches which is of particular
advantage in the multiple alignment setting, as the resulting mean
fields allow straightforward post-processing.

In this paper we
obtain the similarity index
at the maximum a posteriori (MAP) estimates of the rigid-body
transformations and mask parameters because this gives an approximation
to the Kernel Carbo index (\ref{eqkernelcarbo}). We could alternatively
consider a full posterior analysis and work with the posterior distribution.
A similar issue
occurs in Bayesian shape analysis of unlabeled landmark configurations
[\citet{Greemard06}, \citet{Drydenetal07}, \citet{Schmidler07}]
where either a marginal approach (integrating out nuisance parameters)
or a conditional approach (conditioning at the MAP) could be used. We
compared the two approaches
for unlabeled landmarks in other work [\citet{Kenodryd10}]
and the overall performance was similar in the situations considered.
This can be explained by
the similarity of the marginal and conditional
posteriors when a Laplace approximation is accurate (e.g., highly concentrated
posterior distributions for the nuisance parameters).


Finally, as molecules are fuzzy bodies of electronic clouds rather than
discrete sets of atoms, our approach is particularly suited for the
described application. However, as it does not require any
predefined point-by-point correspondence, it could be an approach to
resolve the alignment problem for a fairly broad range of
applications. Examples include matching organs in medical images,
matching objects in images of real-world scenes (e.g., faces) in
photographs or clouds in satellite images.

\section*{Acknowledgments}
The authors would like to thank Jonathan Hirst and James Melville for
motivating discussions about this work, and the Editor and anonymous
referee for many helpful comments.

\begin{supplement}[id-suppA]
\sname{Supplement A}
\stitle{R programs for Bayesian molecule alignment\\}
\slink[doi,text={10. 1214/11-AOAS486SUPPA}]{10.1214/11-AOAS486SUPPA}
\slink[url]{http://lib.stat.cmu.edu/aoas/486/supplementA.zip}
\sdatatype{.zip}
\sdescription{The zip file contains R programs for molecular
alignment using random fields. The main R program is fields8.r which
carries out a Bayesian MCMC procedure.
The programs were written by Irina Czogiel, with some
later edits by Ian Dryden.
There are two options in the program---simulation study
(as in Section \ref{sec4.4}) of the paper, or
comparison of two molecules using steric information
(as in Section \ref{secApplication}).}
\end{supplement}

\begin{supplement}[id-suppB]
\sname{Supplement B}
\stitle{Steroids data}
\slink[doi]{10.1214/11-AOAS486SUPPB}
\slink[url]{http://lib.stat.cmu.edu/aoas/486/supplementB.zip}
\sdatatype{.zip}
\sdescription{The zip file contains the data set of steroids first
analyzed by
\citet{Drydenetal07}.
The data set of $(x,y,z)$ atom co-ordinates and partial charges
was constructed by Jonathan Hirst and James Melville
(School of Chemistry, University of Nottingham).}
\end{supplement}


%

%
\printaddresses

\end{document}